\documentstyle[preprint,aps]{revtex}
\begin{document}
\draft

\title{BIANCHI COSMOLOGICAL MODELS IN THE MINIMUM QUADRATIC
POINCARE GAUGE THEORY OF GRAVITY}
\author{Nguyen Hong Chuong \cite{Chuong} \cite{elm}}
\address{Department of Physics, Syracuse University,
Syracuse, NY 13244-1130, USA}
\author{Nguyen Van Hoang}
\address{Center for Cosmic Physics
and Remote Sensing, 20000 Hochiminh City, Vietnam}
\date{\today}
\maketitle
\begin{abstract}
Within the framework of the minimum quadratic Poincare gauge
theory of gravity in the Riemann-Cartan spacetime the dynamics of
homogeneous anisotropic Bianchi types I-IX spinning-fluid cosmological
models is investigated. A basic equation set for these models is
obtained and analyzed. In particular, exact solutions
for the Bianchi type-I spinning-fluid and Bianchi type-V perfect-fluid
models are found in analytic form.
\end{abstract}
\pacs{SU-GP-93/3-4}

\section{Introduction}
As is well known in the modern theories of fundamental physical
interactions the principle of local gauge symmetry takes the most
important position. On the basis of this principle some unified
gauge theories - the Weinberg-Salam theory, quantum chromodynamics,
Grand Unified theories, etc. - that successfully describe
electromagnetic, weak, and strong interactions of elementary particles,
are constructed and carefully studied. Therefore, it seems natural and
attractive to construct the theory of gravity as the gauge theory of
spacetime symmetries and to study its consequences. During the last 35
years various versions of gauge theory of gravity have been suggested
and investigated. These theories differ one from another by the choice
of the gauge group (Lorentz, Poincare, de Sitter,
metric-affine,  etc.) and gravitational Lagrangian (linear,
quadratic ...). However, the importance of the Poincare symmetry in
particle physics lead one to consider the Poincare gauge theory (PGT)
(see [1-5]) as the most suitable for description of the
gravitational interaction.

In the PGT the gravity is described by two set of gauge fields: the
tetrad fields $\; {\rm h}^{i}{}_{\mu} \>$, and the Lorentz connection
coefficients  $\; {\rm A}^{ik}{}_{\nu} \>$, the spacetime continuum
is represented by a four-dimensional differentiable Riemann-Cartan
manifold $\; {\rm U}_4$.

The corresponding gauge field strength tensors are the torsion:
$${\rm S}^{i}{}_{\mu \nu} = \partial_{[\nu}{\rm h}^{i}{}_{\mu ]} -
{\rm h}_{k[\mu}{\rm A}^{ik}{}_{\nu ]}, $$

\noindent and the curvature:
$$ {\rm F}^{ik}{}_{\mu \nu} = 2 \partial_{[\mu}{\rm A}^{ik}{}_{\nu ]} +
2 {\rm A}^{ij}{}_{[\mu}{\rm A}^{k}{}_{|j| \nu ]}. $$

The sources of the gravitational field in this theory are the canonical
energy-momentum tensor (EMT) $\; {\rm t}_i{}^{\mu}, \> $ and the tetrad
spin current (TSC)  $\; {\rm J}_{ik}{}^{\nu}. \; $

Investigation of gravitating systems in the approximation of homogeneous
isotropic spaces shows, that by including in the gravitational Lagrangian
the terms quadratic in the curvature and torsion tensors the PGT
allows to prevent the appearance of the gravitational singularity
with infinite energy density [3, 5-8].
In connection with avoiding the cosmological singularity
problem, the study of more general homogeneous anisotropic models in
the PGT is of certain interest.

In this article we shall study the dynamics of homogeneous anisotropic
models within the framework of the so-called minimum quadratic (Poincare)
gauge theory of gravity (MQGT) - the simple version of the PGT with the
following gravitational Lagrangian:
$${\rm L}_g = {\rm h} (-\chi + {\it f}_0 {\rm F} + \alpha {\rm F}^2),
\quad {\rm F}_{\mu \nu} = {\rm F}^{\lambda}{}_{\mu \lambda \nu},
\quad {\rm F} = {\rm F}^{\mu}{}_{\mu}, \eqno(1)$$

\noindent where $\; {\rm h} = {\it det}{\rm h}^{i}{}_{\mu}, \>$ {\rm F}
is the scalar curvature of the Riemann-Cartan spacetime ${\rm U}_4$,
$\; {\it f}_0 = (16{\pi}{\rm G})^{-1}, \;$ ({\rm G} is the Newton's
gravitational constant), $\> \chi \>$ is
a cosmological term, and $\; \alpha < 0 \>$ is a dimensionless coefficient
(in the sequel we use all the notations of [9-12] unless otherwise stated).
As was shown in [5, 9] the MQGT satisfies all restrictions following from
the simultaneous consideration of the quantization problem (the theory without
ghosts and tachyons) [13], Birkhoff's
theorem [14], and the problem of cancelling the metric singularity in the
homogeneous isotropic cosmological models [6]. Satisfying the correspondence
principle with General Relativity (GR) for the systems with rather small
energy density, the given theory gives a chance to overcome some difficulties
of GR, which appear on the description of gravitating systems under
extremal conditions (extremely high energy density, spinning matter...).

The paper is organized as follows. In section 2, for spinning-fluid systems
we obtain the gravitational
equations of MQGT in the synchronous frame of references. In section 3, we
derive the equations of motion for Bianchi models in the MQGT. In section
4 and 5 we find analytic solutions of the obtained basic equation set in
the simple cases: Bianchi spinning-fluid type-I and Bianchi perfect-fluid
type-V models, respectively. Some interesting properties of the found
solutions are analyzed in section 6.
\section{MQGT in the synchronous frame of reference.}
By independent variation of Lagrangian (1) with respect to gauge
fields ${\rm h}^i{}_{\mu}$, and ${\rm A}^{ik}{}_{\nu}$ we obtain two field
equations of MQGT in the form:
$$2({\it f}_{0} + 2{\alpha}{\rm F}){\rm F}^{\mu}{}_{i} - ({\it f}_{0} +
{\alpha}{\rm F}){\rm F}{\rm h}^{\mu}{}_{i} + {\chi}{\rm h}^{\mu}{}_{i} =
 -{\rm t}^{\mu}{}_{i}, \eqno(2)$$
$$ 2{\nabla}_{\nu}[({\it f}_{0} + 2{\alpha}{\rm F}){\rm h}_{[i}{}^
{\mu}{\rm h}_{k]}{}^{\nu}] \> = \> -{\rm J}_{ik}{}^{\nu}. \eqno(3)$$

It follows from (2) that $\> {\rm F} = {1\over 2{\it f}_0}({\rm t}^{\mu}_{\mu}
+ 4{\chi}). \>$ Then, it is easily to solve eq. (3):
$${\rm S}^{\lambda}{}_{\mu \nu} = (1 - 4\beta\chi - \beta {\rm t}^{\sigma}_
{\sigma})^{-1}{\bigg(}-{1\over2{\it f}_0}({\rm J}_{\mu \nu}{}^{\lambda} +
{\delta}^{\lambda}_{[\mu}{\rm J}_{\nu ] \sigma}{}^{\sigma}) +
{{\beta} \over 2}{\delta}^{\lambda}_{[\mu} \partial_{\nu ]}{\rm t}
^{\sigma}_{\sigma}{\bigg)}. \eqno(4)$$

By virtue of (4), after some simple transformations (see Ref. [9]) we can
represent (2) in the form of the Einstein's equation with the effective
EMT $\> {\rm T}^{\lambda \mu}_{eff} \>$ constructed from EMT
$\> {\rm t}_{i}^{\mu} \>$ and TSC $\> {\rm J}_{ik}{}^{\nu} \>$ as:
$${\rm G}^{\lambda \mu}(\{ \}) = - {1\over 2{\it f}_0}{\rm T}
^{\lambda \mu}_{eff}, \eqno(5)$$

\noindent where ${\rm G}^{\lambda \mu}_{eff}(\{\})$ is the ordinary
Einstein's tensor and
$${\rm T}^{\lambda \mu}_{eff} \> = \quad (1 - 4\beta\chi - \beta {\rm t}
^{\sigma}_{\sigma})^{-1}{\bigg[}{\rm T}^{\lambda \mu} - {\beta \over 4}{\rm g}
^{\lambda \mu}({\rm t}^{\sigma}_{\sigma} + 4\chi)^2 + \chi {\rm g}^{\lambda
\mu} + {1 \over {2{\it f}_0}}{\Big(}1 - 4\beta\chi -$$
$$- \beta{\rm t}^{\sigma}_{\sigma}{\Big)}^{-1}{\Big[} - 4{\rm J}
^{\lambda \sigma}{}_{[\rho}
 {\rm J}^{\mu \rho}{}_{\sigma ]} - 2 {\rm J}^{\lambda \sigma \rho}
{\rm J}^{\mu}{}_{\sigma \rho} + {\rm J}^{\sigma \rho \lambda}
{\rm J}_{\sigma \rho}{}^{\mu} + {1\over 2}
{\rm g}^{\lambda \mu}(4 {\rm J}_{\nu}{}^{\rho}{}_{[\sigma} {\rm J}
^{\nu \sigma}{}_{\rho ]} + {\rm J}^{\nu \sigma \rho} {\rm J}_{\nu \sigma
\rho}){\Big]}{\bigg]} +$$
$$+ 2{\it f}_0{\Big[}{\tilde {\nabla}}^{\mu}{\partial}^{\lambda}{\rm M} -
{1\over 2}{\partial}^{\mu}{\partial}^{\lambda}{\rm M}
- {\rm g}^{\lambda \mu}({\tilde {\nabla}}_{\sigma}{\partial}^{\sigma}{\rm M}
+ {1\over 4}{\partial}_{\sigma}{\rm M}{\partial}^{\sigma}{\rm M}){\Big]},
 \eqno(6)$$

\noindent where $\> {\rm M} = {\rm ln}{|1 - 4\beta\chi - \beta{\rm t}^{\sigma}
_{\sigma}|}$,
$\> \beta = - {\alpha \over {\it f}_O^2} > 0$, $\> {\rm T}^{\lambda \mu} =
{\rm t}^{\lambda \mu} + ({\nabla}_{\nu} + 2{\rm S}_{\nu})({\rm J}^{\mu
\lambda \nu} + {\rm J}^{\lambda \nu \mu} + {\rm J}^{\mu \nu \lambda})$, $\>
{\rm S}_{\nu} = {\rm S}^{\sigma}{}_{\sigma \nu}$, $\> {\nabla}_{\nu} \>$ and
$\> {\tilde \nabla}_{\nu} \>$ denote covariant derivatives calculated by means
of
the total connection $\> {\Gamma}^{\lambda}{}_{\mu \tau} \>$ and the
Christoffel
symbols $\> {\gamma}^{\lambda}{}_{\mu \tau} \>$ respectively.

As the source of gravitational fields we consider the perfect fluid with
angular spin momentum (spinning fluid). In [15-18] a variational
formalism for relativistic dynamics of spinning fluid was developed
in the Riemann-Cartan spacetime. In the simplest spinning  fluid model
the canonical EMT $\> {\rm t}^{\lambda \mu} \>$ and TSC $\> {\rm J}
_{\mu \nu}{}^{\lambda} \>$ take the following form:
$${\rm t}^{\lambda \mu} \> = \> (\rho + {\rm p}){\rm u}^{\lambda}
{\rm u}^{\mu} - {\rm p}{\rm g}^{\lambda \mu} + {\rm j}^{\lambda}{}
_{\tau}{\rm u}^{\mu}{\rm u}^{\sigma}{\nabla}_{\sigma}{\rm u}^{\tau},
\eqno(7)$$
$${\rm J}_{\mu \nu}{}^{\lambda} \> = \> {1\over 2}{\rm j}_{\mu \nu}
{\rm u}^{\lambda}, \eqno(8)$$

\noindent where $\> {\rho} \>$ is the invariant energy density, $\> {\rm p} \>$
ia the pressure, $\> {\rm u}^{\mu} \>$ is the 4-velocity vector,
$\> {\rm j}_{\mu \nu} = - {\rm j}_{\nu \mu} \>$ is the spin density
tensor. By using the relations:
$\> {\rm j}_{\mu \nu}{\rm u}^{\nu} = 0\>$ from (7)
we get $\> {\rm t}^{\sigma}_{\sigma} = {\rho} - 3{\rm p} \>$. Then,
the torsion tensor (4) can be rewritten as
$${\rm S}^{\lambda}{}_{\mu \nu} = (1 - 4\beta\chi - \beta {\rm t}
^{\sigma}_{\sigma})^{-1}{\bigg(} - {1\over 4{\it f}_0}{\rm j}
_{\mu \nu}{\rm u}^{\lambda} + {\beta \over 2}{\delta}^{\lambda}
_{[ \mu}{\partial}_{\nu ]}{\rm t}^{\sigma}_{\sigma}{\bigg)}. \eqno(9)$$

In view of (7)-(9) and the equations of rotational motion (see [15,16])
the effective EMT $\> {\rm T}^{\lambda \mu}_{eff} \>$ can be transformed
to the form:
$${\rm T}^{\lambda \mu}_{eff} \> = \> (1 - 4\beta\chi - \beta {\rm t}
^{\sigma}_{\sigma})^{-1}
{\bigg[}{\bigg(}{\rho} + {\rm p} - {{{\rm j}^2}\over {4{\it f}_0}}(1 - 4
\beta\chi - \beta{\rm t}^{\sigma}_{\sigma})^{-1}{\bigg)}{\rm u}^{\lambda}
{\rm u}^{\mu} - {\rm g}^{\lambda \mu}{\bigg(}{\rm p} - {\chi} +$$
$$+  {\beta \over 4}({\rm t}^{\sigma}_{\sigma} + 4\chi)^2 -
{{{\rm j}^2} \over {8{\it f}_0}}(1 - 4\beta\chi - \beta{\rm t}^{\sigma}
_{\sigma})^{-1}{\bigg)} + {\rm j}^{\nu ( \lambda}{\rm u}^{\mu )}{\partial}
_{\nu}{\rm M} - {\rm u}^{\nu}{\rm u}^{\tau}{\rm u}^{( \mu}{\tilde \nabla}
_{\nu}{\rm j}^{\lambda )}{}_{\tau} -$$
$$-  {\tilde \nabla}_{\nu}({\rm j}^{\nu ( \lambda}{\rm u}^{\mu )}){\bigg]} +
2{\it f}_{0}{\Big[}{\tilde \nabla}^{\mu}{\partial}^{\lambda}{\rm M}
- {1 \over 2}{\partial}^{\mu}{\rm M}{\partial}^{\lambda}{\rm M}
- {\rm g}^{\lambda \mu}
({\tilde \nabla}_{\nu}{\partial}^{\nu}{\rm M} + {1 \over 4}{\partial}
_{\nu}{\rm M}{\partial}^{\nu}{\rm M}){\Big]}, \eqno(10)$$

\noindent where $\> {\rm j}^{2} = {1 \over 2}{\rm j}_{\mu \nu}
{\rm j}^{\mu \nu}.$

Following [19] we rewrite the MQGT gravitational equations in the synchronous
frame of reference, in which the spacetime metric takes the form:
$${\rm ds}^2 \> = \> {\rm d}{\it t}^2 - {\gamma}_{\alpha \beta}{\rm dx}
^{\alpha}{\rm dx}^{\beta}, \eqno(11)$$

\noindent where $\; {\it t} \;$ is a synchronous time, $\> {\rm x}^{\alpha}
- \>$ spatial coordinates (first greek letters ${\alpha}, {\beta}$ ...
pass values 1, 2, 3). As in [19] let us introduce the notation:
$$ {\cal H}_{\alpha \beta} \; = \; {\dot \gamma}_{\alpha \beta} \eqno(12)$$

\noindent (a dot denotes differentiation with respect to the synchronous
time $\;{\it t} \>$).
The quantities $\> {\cal H}_{\alpha \beta} \>$ make up a 3-dimensional tensor,
raising or lowering of one's indices, and so the covariant differentiation is
calculated by the 3-metric $\; {\gamma}_{\alpha \beta}. \>$ In terms of
$\; {\gamma}_{\alpha \beta} \>$ and $\; {\cal H}_{\alpha \beta} \;$
the Einstein tensor's components
can be expressed as:
$${\rm G}^0_0 \; = \; {1\over 8}{\cal H}^{\alpha}_{\beta}{\cal H}^{\beta}
_{\alpha} - {1\over 8}{{\dot \gamma} \over {\gamma}}{\cal H}^{\alpha}
_{\alpha} - {1 \over 2}{\cal P}^{\alpha}_{\alpha},$$
$${\rm G}^0_{\alpha} \; = \; {1 \over 2}({\tilde \nabla}_{\alpha}{\cal H}
^{\beta}_{\beta} - {\tilde \nabla}_{\beta}{\cal H}^{\beta}_{\alpha}),$$
$${\rm G}^{\beta}_{\alpha} \; = \; {1 \over 2}{\dot {\cal H}}^{\beta}
_{\alpha} + {1 \over 4}{{\dot {\gamma}} \over {\gamma}}{\cal H}^{\beta}
_{\alpha} + {\cal P}^{\beta}_{\alpha} - {1 \over 2}{\delta}^{\beta}_{\alpha}
{\big(}{\dot {\cal H}}^{\delta}_{\delta} + {1 \over 4}{{\dot {\gamma}}
 \over {\gamma}}
{\cal H}^{\delta}_{\delta} + {1 \over 4}{\cal H}^{\gamma}_{\delta}{\cal H}
^{\delta}_{\gamma} + {\cal P}^{\delta}_{\delta}{\big)}, \eqno(13)$$

\noindent where  $\> {\gamma} = {\it det}({\gamma}_{\alpha \beta}),
\; {\cal P}
^{\beta}_{\alpha} \;$  is  a 3-dimensional Ricci tensor constructed
from $\> {\gamma}_{\alpha \beta}. \>$ Concerning the source of gravitational
field we will consider only the case of rest matter, when $\> {\rm u}^0
= 1, \; {\rm u}^{\alpha} = 0, \>$ (hence we have $\> {\rm j}_{\mu 0} = 0). \>$
Using these relationships we obtain the following expressions for the
components of $\> {\rm T}^{\lambda \mu}_{eff} \>$ in the synchronous
frame of reference:
$${\rm T}^0_{0(eff)} \; = \; (1 - 4\beta\chi - \beta {\rm t}^{\sigma}
_{\sigma})^{-1}{\big[}{\rho} + \chi - {\beta \over 4}({\rm t}^{\sigma}
_{\sigma} + 4\chi)^2 - {1 \over 8{\it f}_0}{\rm j}^2(1 - 4\beta\chi -
\beta{\rm t}^{\sigma}_{\sigma})^{-1}{\big]} - $$
$$ - \> 2{\it f}_0({1\over 2}
{\dot {\rm M}}{\cal H}^{\alpha}_{\alpha} - {3 \over 4}{\dot {\rm M}}^2), $$
$${\rm T}^0_{\alpha (eff)} \; = \; - {1 \over 2}(1 - 4\beta\chi
- \beta {\rm t}^{\sigma}_{\sigma})^{-1}{\tilde \nabla}_{\beta}
{\rm j}^{\beta}{}_{\alpha},$$
$${\rm T}^{\beta}_{\alpha (eff)} \; = \; - (1 - 4\beta\chi - \beta
{\rm t}^{\sigma}_{\sigma})^{-1}{\bigg[}{\big[}{\rm p} - \chi +
{\beta \over 4}
({\rm t}^{\sigma}_{\sigma} + 4\chi )^2 - {1 \over 8{\it f}_0}{\rm j}^2
(1 - 4\beta\chi - \beta {\rm t}^{\sigma}_{\sigma})^{-1}{\big]}{\delta}
^{\alpha}_{\beta} -$$
$$- {1 \over 4}({\cal H}^{\gamma}_{\alpha}{\rm j}^{\beta}{}_{\gamma}
+ {\cal H}^{\beta}_{\gamma}{\rm j}^{\gamma}{}_{\alpha}){\bigg]}
+ {2{\it f}_{0}{\Big[}{1 \over 2}{\dot {\rm M}}{\cal H}^{\beta}_{\alpha}
+{\big(}- {\ddot {\rm M}} + {1 \over 2}{\dot {\rm M}}{\cal H}^{\gamma}_{\gamma}
+ {1 \over 4}{\dot {\rm M}}^2{\big)}{\delta}^{\beta}_{\alpha}{\Big]}}.
\eqno(14)$$

Note that covariant differentiation $\> {\tilde \nabla} \>$ of the quantity
$\> {\rm j}^{\beta}{}_{\alpha} \>$ in the expression of $\; {\rm T}
^0_{\alpha (eff)} \;$ is calculated by the 3-metric $\> {\gamma}_
{\alpha \beta}. \>$
\section{Bianchi models in the MQGT: a basic equation set}
Let us proceed to derive the equations of motion for homogeneous
anisotropic (Bianchi) cosmological models. Homogeneity of the 3-space
implies invariance of the spatial metric:
$${\rm dl}^2 \; = \; {\gamma}_{\alpha \beta}{\rm dx}^{\alpha}
{\rm dx}^{\beta}$$

\noindent at each fixed moment of time $\; {\it t} \;$ with respect to the
3-dimensional group of motion $\> G_3 \>$ (see [19-21]). So that
local characters
of homogeneous anisotropic models are determined fully by types of
 the corresponding
motion's group G, which was classified by Bianchi with respect to
independent sets of the structure contants $\; {\rm C}^{c}{}_{ab} \>$.
Here and later on, the first latin letters $\; {\rm a, b, c...} \;$
take the values $ \; {\rm 1, 2, 3}  \;$ and number
of the basis's vectors.

Let $\; {\rm X}_{a} \> = \> {\rm e}^{\alpha}{}_{a}{{\partial} \over
{\partial}{\rm x}^{\alpha}}, \;$ ({\rm a = 1, 2, 3}) be generators
of G, that satisfy the following commutation relations:
$${\rm [X}_{a}, {\rm X}_{b}{\rm ]} \; = \; {\rm C}^{c}{}_{ab}{\rm X}
_{c},  \eqno(15)$$

\noindent where $\; {\rm e}^{\alpha}{}_{a} \;$ are the basis's vectors of a
3-dimentional space with the metric $\> {\gamma}_{\alpha \beta}. \>$
As we see from (15) the structure constants are antisymmetric in their
lower indices: $\; {\rm C}^c{}_{ab} = - {\rm C}^c{}_{ba}, \>$ and
satisfy the Jacobi identity:
$${\rm C}^e{}_{ab}{\rm C}^d{}_{ec} + {\rm C}^e{}_{bc}{\rm C}^d{}_{ea}
+ {\rm C}^e{}_{ca}{\rm C}^d_{eb} \; = \; 0.$$

We also introduce basis's vectors $\> {\rm e}^{a}_{\alpha} \>$
that satisfy the relationships:
$${\rm e}^{a}_{\alpha}{\rm e}^{\alpha}_{b} \; = \;
{\delta}^{a}_{b}; \qquad {\rm e}^{\alpha}_{a}{\rm e}^{a}_{\beta}
\; = \; {\delta}^{\alpha}_{\beta}. \eqno(16)$$

Expanding the spatial metric with respect to the basis $\> ({\rm e}
^{\alpha}{}_{a}) \>$ we obtain the following time-dependent functions:
$${\eta}_{ab}({\it t}) \; = \; {\gamma}_{\alpha \beta}{\rm e}^{\alpha}{}_{a}
{\rm e}^{\beta}{}_{b}. \eqno(17)$$

The Bianchi classification of homogeneous spaces reduces to determining all
nonequivalent sets of structure constants. In place of the three-index
constants we introduce a set of two-index quantities, defined by the
dual transformation:
 $\; {\rm C}^c{}_{ab} \> = \> {\varepsilon}_{abd}{\rm C}^{dc}, \;$ (
${\varepsilon}_{abc} \> = \> {\varepsilon}^{abc} \;$ is the unit antisymmetric
symbol with $\; {\varepsilon}_{123} \> = \> 1. \>$ Following [22], we decompose
nonsymmetric $\; {\rm C}^{ab} \>$ into a symmetric tensor $\; {\rm n}^{ab} \>$
and a vector $\; {\rm a}_c \>$ as:
$${\rm C}^{ab} \; = \; {\rm n}^{ab} + {\varepsilon}^{abc}{\rm a}_c. $$

It is easy to diagonalize the symmetric tensor $\; {\rm n}^{ab} \>$; let
$\; {\rm n}_1, \> {\rm n}_2, \> {\rm n}_3 \>$ be its eigenvalues.
It then follows from the Jacobi identity that if $\; {\rm a}_a \>$ exits it
is an eigenvector of $\; {\rm n}^{ab} \>$ corresponding to a zero eigenvalue:
$${\rm n}^{ab}{\rm a}_a \> = \> 0.$$

Without loss of generality the vector $\; {\rm a}_a \>$ can be written as
$\; {\rm a}_a \> = \> ({\rm a}, 0, 0) \>$ with $\; {\rm n}_1{\rm a} = 0. \>$
Then, we get either $\; {\rm n}_1 \>$ or $\; {\rm a} \>$ is equal to zero.
Bianchi models with $\; {\rm a} = 0 \>$ were defined by Ellis and MacCallum as
A-class models,  also known as tensor models. Bianchi models for
which $\; {\rm n}_1 \>$ is set equal to zero are known as B-class models -
vector models. By rescaling the operators $\; {\rm X}_1, \> {\rm X}_2,
\> {\rm X}_3 \> $ one can set the eigenvalues $\; {\rm n}_1, \> {\rm n}_2,
\> {\rm n}_3 \>$ to be equal to $\> 0, {\pm}1. \>$ The classification of
these models is given in the tables A, and B, respectively (see [19-20]):

\begin{center}
{\sl Table A: Classification of A-class models} {\rm (a = 0)}

\bigskip
\begin{tabular}{|c|c|c|c|} \hline
$\qquad Type \qquad$ & $\quad {\rm n}_1 \quad$ & $\quad {\rm n}_2 \quad$ &
$\quad {\rm n}_3 \quad$
\\ \hline
I & 0 & 0 & 0 \\  \hline
II & 1 & 0 & 0 \\ \hline
$VII_0$ & 1 & 1 & 0 \\ \hline
$VI_0$ & 1 & $\quad -1 \quad$ & 0 \\  \hline
IX & 1 & 1 & 1 \\    \hline
VIII & 1 & 1 & -1 \\ \hline
\end{tabular}

\bigskip

{\sl Table B: Classification of B-class models $\; (n_1 = 0)$}

\bigskip

\begin{tabular}{|c|c|c|c|} \hline
$\qquad Type  \qquad$ & $\quad {\rm a} \quad$ & $\quad {\rm n}_2 \quad$ &
$\quad {\rm n}_3 \quad$
\\ \hline
V & 1 & 0 & 0 \\ \hline
IV & 1 & 0 & 1 \\ \hline
III & 1 & -1 & 1 \\ \hline
$VI_h$ & $\quad {\rm a} > 0 \quad$ & $\quad -1 \quad$ & 1 \\ \hline
$VII_h$ & ${\rm a} > 0$ & 1 & 1 \\ \hline
\end{tabular}
\end{center}

\bigskip
By virtue of (12) and (15)-(17) the expansions of ${\cal H}^{\beta}_{\alpha},
\; {\cal P}^{\beta}_{\alpha}, \; ({\tilde \nabla}_{\alpha}{\cal H}^{\beta}
_{\beta} - {\tilde \nabla}_{\beta}{\cal H}^{\beta}_{\alpha}),\;$ and
${\tilde \nabla}_{\beta}{\rm j}^{\beta}_{\alpha}$ with respect to
$\; \{ {\rm e}^{\alpha}{}_{a} \} \;$ lead to the following expressions
(see [19]):
$${\cal H}^a_b \: = \; {\cal H}^{\beta}_{\alpha}{\rm e}^{\alpha}{}_{a}
{\rm e}^b{}_{\beta} \; = \; {\eta}^{ac}{\dot \eta}_{bc}, \quad
{\cal H}^a_a \; = \; {\cal H}^{\alpha}_{\alpha}, \eqno(18)$$
$${\cal P}^b_a \; = \; {1 \over 2{\eta}}{\Big[}2{\rm C}^{bd}{\rm C}_{ad}
+ {\rm C}^{db}{\rm C}_{ad} + {\rm C}^{bd}{\rm C}_{da} - {\rm C}^d{}_d
({\rm C}^b{}_a +{\rm C}_a{}^b) + {\delta}^b_a{\big[}({\rm C}^d{}_d)^2
- 2{\rm C}^{cd}{\rm C}_{cd}{\big]}{\Big]}, \eqno(19)$$
$${\rm e}^{\alpha}{}_{a}({\tilde \nabla}_{\alpha}{\cal H}^{\beta}_{\beta}
- {\tilde \nabla}_{\beta}{\cal H}^{\beta}_{\alpha}) \; = \;
{\cal H}^{c}_{b}({\rm C}^b{}_{ca} - {\delta}^b_a{\rm C}^d{}_{dc}), \eqno(20)$$
$${\rm e}^{\alpha}{}_{a}({\tilde \nabla}_{\beta}{\rm j}^{\beta}{}_{\alpha})
\; = \; {1 \over 2}{\rm j}^b{}_c{\rm C}_{ab}{}^c - {\rm j}^b{}_a{\rm
C}^c{}_{bc}
, \eqno(21)$$

\noindent where $\; {\eta} \> = \> {\it det}({\eta}_{ab}),\;
{\rm j}^b{}_c \> = \> {\rm j}^{\beta}{}_{\gamma}{\rm e}^{\gamma}
{}_{c}{\rm e}^b{}_{\beta}.\;$ and, in accordance with the general rule:
$${\rm C}^b_a \; = \; {\eta}_{ac}{\rm C}^{cb}; \quad {\rm C}_{ab} \; =
\; {\eta}_{ac}{\eta}_{bd}{\rm C}^{cd}.$$

Substituting (17)-(21) to the expressions (13), (14) we find the following
expansions of the components of $\; {\rm G}^{\lambda \mu} \;$  and $\;
{\rm T}^{\lambda \mu}_{eff} \;$ with respect to the basis
$\; \{ {\rm e}^a{}_{\alpha} \}:$
$${\rm G}^0_0 \; = \; {1 \over 8}{\Big(}{\cal H}^a_b{\cal H}^b_a -
{{\dot {\eta}} \over {\eta}}{\cal H}^a_a - 4{\cal P}^a_a{\Big)},$$
$${\rm G}^0_a \; = \; {1 \over 2}{\cal H}^c_b({\rm C}^b_{ca} - {\delta}
^b_a{\rm C}^d_{dc}),$$
$${\rm G}^b_a \; = \; {1 \over 2}{\dot {\cal H}}^b_a + {1 \over 4}{{\dot
{\eta}} \over {\eta}}{\cal H}^b_a + {\cal P}^b_a - {1 \over 2}{\delta}
^b_a{\Big(}{\dot {\cal H}}^c_c + {1 \over 4}{{\dot {\eta}} \over {\eta}}{\cal
H}
^c_c + {1 \over 4}{\cal H}^c_d{\cal H}^d_c + {\cal P}^c_c{\Big)}, \eqno(22)$$
$${\rm T}^0_{0(eff)} \; = \; (1 - 4\beta\chi - \beta{\rm t}^{\sigma}_{\sigma})
^{-1}{\Big[}{\rho} + {\chi} - {{\beta} \over 4}({\rm t}^{\sigma}_{\sigma}
+ 4\chi)^2{\Big]}
- {{\rm j}^2 \over {8{\it f}_0}}(1 - 4\beta\chi - \beta{\rm t}^{\sigma}
_{\sigma})^{-2} - 2{\it f}_0({1 \over 2}{\dot {\rm M}}{\cal H}^a_a +
{3 \over 4}{\dot {\rm M}}^2), $$
$${\rm T}^0_{a(eff)} \; = \; - {1 \over 2}(1 - 4\beta\chi - \beta{\rm t}
^{\sigma}_{\sigma})^{-1}{\Big(}{1 \over 2}{\rm j}^b{}_c{\rm C}_{ab}{}^c
- {\rm j}^b{}_a{\rm C}^c{}_{bc}{\Big)},$$
$${\rm T}^b_{a(eff)} \; = \; {\bigg[}-(1 - 4\beta\chi - \beta{\rm t}
^{\sigma}_{\sigma})^{-1}{\Big[}{\rm p} - {\chi} + {{\beta} \over 4}
({\rm t}^{\sigma}_{\sigma} + 4\chi)^2{\Big]} +{{\rm j}^2 \over {8{\it
f}_0}}(1 - 4\beta\chi - \beta{\rm t}^{\sigma}_{\sigma})
^{-2}{\bigg]}{\delta}^{b}_{a} -$$
$$- {1 \over 4}(1 - 4\beta\chi - {\beta}{\rm t}^{\sigma}_{\sigma})^{-1}
{\Big(}{\rm j}_c{}^b{\cal H}^c_a + {\rm j}^c{}_a{\cal H}^b_c{\Big)} -
2{\it f}_0{\Big[}{\Big(}{\ddot {\rm M}} + {1 \over 2}{\dot {\rm M}}
{\cal H}^c_c + {1 \over 4}{\dot {\rm M}}^2{\Big)}{\delta}^b_a -
{1 \over 2}{\dot {\rm M}}{\cal H}^{b}_{a}{\Big]}. \eqno(23)$$

Thus, we obtain the MQGT basic equation set describing the dynamics
of Bianchi (types I-IX) spinning-fluid cosmological models, written
in the basis $\; \{ {\rm e}^{\alpha}{}_a \}, \;$ in the
following form:
$${\rm G}^0_0 \; = \; - {1 \over {2{\it f}_0}}{\rm T}^0_{0(eff)}, \quad
{\rm G}^0_{a} \; = \; {1 \over {2{\it f}_0}}{\rm T}^0_{a(eff)}, \quad
{\rm G}^b_a \; = \; {1 \over {2{\it f}_0}}{\rm T}^b_{a(eff)}, \eqno(24)$$

\noindent where the corresponding compononents of $\; {\rm G}^{\lambda \mu}
\;$ and
$\; {\rm T}^{\lambda \mu}_{eff} \;$ are determined by the expressions
(22)-(23), which contain only the time-dependent functions.

Note that these equations
reduce to the corresponding equations of the Einstein-Cartan theory in the
case when $\; {\beta} \> = \> 0, \;$ and coincide with the ordinary
Einstein's equations of GR in the case when $\; {\beta}
\> = \> 0 \;$ and $\; {\rm j}_{\mu \nu} \> = \> 0.$

In the following sections we will study Bianchi models
of types I and V, which are the direct anisotropic generalization of the
flat and the open isotropic Friedmann-Robertson-Walker models
respectively.
\section{Bianchi type-I spinning-fluid models: analytic solutions}
We now study the dynamics of the simplest case of homogeneous
anisotropic - Bianchi type-I - models, when all of the structure
constants $\; {\rm C}^c_{ab} \> = \> 0 \;$ (see also [12]). Then,
 the metric of 3-space can be written in the diagonal form:
$${\eta}_{ab} \; = \; {\it diag}{\Big(}{\rm r}^2_1({\it t}), \;
{\rm r}^2_2({\it t}), \; {\rm r}^2_3({\it t}){\Big)}. \eqno(25)$$

In this case we note that under the presence of spin momentum $\;
{\rm j}^a{}_b
\not= 0 \>$ the 3-space will be either isotropic or axially symmetric.
In fact, for $\; a \not= b \>$ we find from (22)  $ \; {\rm G}^a_b = 0 \;$
and from (23)
$${\rm T}^b_{a(eff)} \; = \; - {1 \over 4}(1 - 4\beta\chi -
\beta{\rm t}^{\sigma}_{\sigma})^{-1}{\Big(}{\rm j}_c{}^b{\cal H}^c_a + {\rm j}
^c{}_a{\cal H}^b_c{\Big)} \; =$$
$$= \; - {1 \over 4}(1 - 4\beta\chi -\beta{\rm t}^{\sigma}_{\sigma})^{-1}
{\Big(}{\cal H}^{\underline a}_{\underline a} - {\cal H}^{\underline b}
_{\underline b}{\Big)}{\rm j}_{{\underline a} {\underline b}}{\eta}
^{{\underline a} {\underline b}}$$

\noindent (there is no sum on indexes {\underline a}, {\underline b} here).
It follows that
$${\Big(}{{\dot {\rm r}_{\underline a}} \over {{\rm r}_{\underline a}}} -
{{\dot {\rm r}_{\underline b}} \over {{\rm r}_{\underline b}}}{\Big)}
{\rm j}_{{\underline a}{\underline b}} \; = \; 0 \quad ({\underline a},
{\underline b} = 1, 2, 3;
\;{\underline a} \not= {\underline b}). \eqno(26)$$

\noindent In the spinless matter case $\; ({\rm j}_{ab} = 0) \;$ relation (19)
is automatically satisfied. If at least two of the three components of
$\; {\rm j}_{ab} \;$ are not equal to zero, then (26) implies
$\; {{\dot {\rm r}_1} \over {{\rm r}_1}} = {{\dot {\rm r}_2} \over
{{\rm r}_2}} = {{\dot {\rm r}_3} \over {{\rm r}_3}}, \;$ which leads to
the isotropic 3-space: $\; {\rm r}_1 = {\rm r}_2 = {\rm r}_3. \;$
If only one of the three components of $\; {\rm j}
_{ab} \;$ is nonzero, for example $\; {\rm j}_{12}, \;$ it follows that
the 3-space will be axially symmetric $\; {\rm r}_1 = {\rm r}_2 \not=
{\rm r}_3 \;$

By analogy with GR we introduce new variables:
$\; {\rm r} \> = \> {({\rm r}_1{\rm r}_2{\rm r}_3)}^{1/3}, \;
{\rm h} \> = \> ({\dot {\rm r}}/{\rm r}), \; {\rm h}_k \> = \>
({\dot {\rm r}}_{k}/{\rm r}_{k}) \; ({\rm k} \> = \> 1, 2. 3). \>$
Let us consider the general case, when $\; {\rm t}^{\sigma}_{\sigma}
 \> = \> \rho - 3{\rm p} \> \not= \> 0.$
\footnote{The special case of radiation $\; {\rm t}^{\sigma}_{\sigma}
= {\rho} - 3{\rm p} = 0 \;$ was carefully studied in [11].}
Then, it is easy to find
$${\cal H}^b_a \; = \; 2 \> {\it diag}(\> {\rm h}_1, \> {\rm h}_2,
\> {\rm h}_3 \>); \quad {\eta} \> = \> {\rm r}^6; \quad
{\cal P}^b_a \> = \> 0. \eqno(27)$$

By using (27) we can rewrite (22) and (23) as
$${\rm G}^0_0 \; = \; {1 \over 2}{\sum}_{k=1}^3 {\rm h}^2_k - {9 \over 2}
{\rm h}^2,$$
$${\rm G}^{\underline a}_{\underline a} \; = \; {\dot {\rm h}_a}
- 3{\dot {\rm h}} + 3{\rm h}{\rm h}_a -  {1 \over 2}{\sum}_{k=1}^3 {\rm h}^2_k
- {9 \over 2}{\rm h}^2 ,$$
$${\rm T}^0_{0(eff)} \; = \; (1 - 4\beta\chi - \beta{\rm t}^{\sigma}_{\sigma})
^{-1}{\Big[}{\rho} + {\chi} - {{\beta} \over 4}({\rm t}^{\sigma}_{\sigma}
 + 4\chi)^2{\Big]} -$$
$$-  {{\rm j}^2 \over {8{\it f}_0}}(1 - 4\beta\chi -
\beta{\rm t}^{\sigma}_{\sigma})^{-2} - 6{\it f}_0({1 \over 2}{\dot
{\rm M}}{\rm h} + {1 \over 4}{\dot {\rm M}}^2), $$
$${\rm T}^{\underline a}_{{\underline a}(eff)} \; = \; - (1 - 4\beta\chi
- \beta{\rm t}^{\sigma}_{\sigma})^{-1}{\Big[}{\rm p} - {\chi} +
{{\beta} \over 4}({\rm t}^{\sigma}_{\sigma} + 4\chi)^2{\Big]} +$$
$$+ {{\rm j}^2 \over {8{\it f}_0}}(1 - 4\beta\chi - \beta{\rm t}^{\sigma}
_{\sigma})^{-2} - 2{\it f}_0{\Big[}{\ddot {\rm M}} + {\dot {\rm M}}
(3{\rm h} - {\rm h}_a) + {1 \over 4}{\dot {\rm M}}^2{\Big]}. \eqno(28)$$

After some manipulations from (24) and (28) we obtain
(for the case of $\> \chi = 0 \>$ see also [10])
$${\rm h}_k \; = \; {\rm h} + s_k{\rm r}^{-2}(1 - 4\beta\chi -
\beta{\rm t}^{\sigma}_{\sigma})^{-1}; \quad {\sum}_{k=1}^{3}s_k \>
= \> 0, \eqno(29)$$
$${\Big[}{\rm h} - {\beta \over 2}(1 - 4\beta\chi -
\beta{\rm t}^{\sigma}_{\sigma})^{-1}({\rm t}^{\sigma}_{\sigma})^{.}
{\Big]}^2 \; = \; (1 - 4\beta\chi -
\beta{\rm t}^{\sigma}_{\sigma})^{-1}{\bigg[}{1 \over {6{\it f}_0}}
{\Big(}{\rho} + {\chi} - {\beta \over 4}({\rm t}^{\sigma}_{\sigma}
+ 4\chi)^2{\Big)} +$$
$$+ (1 - 4\beta\chi - \beta{\rm t}^{\sigma}_{\sigma})^{-1}
{\rm r}^{-6}{\bigg(}{\sl l}^2  - {{\rm j}^2 \over {48{\it f}_0^2}}
{\bigg)}{\bigg]}, \eqno(30)$$
$${\dot {\rm h}} + 3{\rm h}^2 \; = \; (1 - 4\beta\chi - \beta{\rm t}
^{\sigma}_{\sigma})^{-1}{\bigg[}{1 \over 4{\it f}_0}{\Big(}{\rho} -
{\rm p} + 2\chi - {\beta \over 2}({\rm t}^{\sigma}_{\sigma}
+ 4\chi)^2{\Big)} +$$
$$+ {\beta \over 2}{\Big(}({\rm t}^{\sigma}_{\sigma})^{..} + 5{\rm h}
({\rm t}^{\sigma}_{\sigma})^{.}{\Big)}{\bigg]}, \eqno(31)$$

\noindent where $\; s_k \> (k = 1, 2, 3) \;$ are integration constants and
$\; {\sl l}^2 = {1 \over 6}{\sum}_{k=1}^{3}{s_k^2} \;$ is a measure of the
anisotropy.

In the case considered, the spin conservation law (by using
the equation of rotation of fluid in [15]) takes the following simple form:
$$({\rm j})^{.} + 3{\rm h}{\rm j} \; = \; 0. \eqno(32)$$

and "the energy-momentum conservation law" $\; {\tilde \nabla}_{\mu}
{\rm T}^{\lambda \mu}_{eff} = 0 \;$ yields
$${\dot {\rho}} + 3{\rm h}(\rho + {\rm p}) \; = \; 0. \eqno(33)$$

{}From these conservation laws we obtain
$${\rm r} \; = \; exp{\bigg(} - {1 \over 3}{\int}{{d{\rho}} \over {{\rho}
 + {\rm p}({\rho})}}{\bigg)}; \quad {\rm j}{\rm r}^3 \> = \> {\rm J}_0
\> = \> const. \eqno(34)$$

Assuming $\; {\rho} = {\rho}({\rm r}), \; {\rm j} = {\rm j}({\rm r}) \;$
in accordance with (34), it is easy to find the solution of (30) in an
analytic form:
$${\it t - t_0} \; = \; \int_{{\rm r}_0}^{\rm r} {{\Phi}_1({\rm r}) \over
{\Phi}_2^{1/2}({\rm r})}{\rm dr}, \eqno(35)$$

\noindent where
$${\Phi}_1({\rm r}) \; = \; {3 \over {\rm r}}{\bigg[}1 - 4\beta\chi -
\beta{\rm t}^{\sigma}_{\sigma} +{3\beta \over 2}(\rho +
{\rm p}){\Big(} 1 - 3{{\rm dp} \over {d{\rho}}}{\Big)}{\bigg]},$$
$${\Phi}_2({\rm r}) \; = \; {3 \over {2{\it f}_0}}(1 - 4\beta\chi - \beta{\rm
t}
^{\sigma}_{\sigma}){\Big[}{\rho} + \chi - {\beta \over 4}{\big(}{\rm t}
^{\sigma}_{\sigma} + 4\chi{\big)}^2{\Big]} + {{\sl A} \over {{\rm r}^6}},$$

\noindent and $\; {\sl A} \> = \>9{\Big(} {\sl l}^2 - {{\rm J}_0^2 \over
{48{\it f}_0^2}}{\Big)}.\;$
It follows from (29) that
$${\rm r}_k \; = \; {\rm r} \> exp{\Big(}s_k{\int}{\rm r}^{-3}(1 -
4\beta\chi - \beta{\rm t}^{\sigma}_{\sigma})^{-1}d{\it t}{\Big)}. \eqno(36)$$

Thus, under the given equation of state $\; {\rm p} = {\rm p}(\rho) \;$ the
analytic solutions (34)-(36) adequately describe the dynamics of the
Bianchi spinning-fluid models. Note that the solution (35) makes sense only
if $\; {\Phi}_2({\rm r}) \geq 0. \;$ For the case of vanishing
cosmological term $\; \chi = 0 \;$ the study of these solutions (for more
details see Ref [23]) shows that we can construct Bianchi type-1
cosmological models, which are regular in the metric and in the torsion,
by fulfilling the condition $\; A < 0 \;$ (i.e. when the spin effect
dominates over the effect of anisotropy). For the case of $\; \chi \not= 0 \;$
and the linear equation of state:
$${\rm p} \; = \; {{1 - \gamma} \over 3}{\rho}, \quad 0 < \gamma \leq 1,
\eqno(37)$$

\noindent a complete qualitative analysis of properties of the obtained
solutions is perfomed in [12] by using methods of qualitative theory of
dynamic systems [24].
\section{Bianchi type-V perfect-fluid models: analytic solutions.}
In this section we restrict ourselves to the spinless matter
case of the perfect fluid, when $\; {\rm j}_{\mu \nu} \> = \> 0.$
As was shown in [25] in the general case of Bianchi models (except for
type-I) the metrics of anisotropic models can not
be diagonalized in the presence of spin momentum.

Taking $\; ({\rm x}_1, \> {\rm x}_2, \> {\rm x}_3) \;$ as local coordinates
we can write the interval of Bianchi type-V models in the diagonal form [25,
26]:
$${\rm ds}^2 \; = \; d{\it t}^2 - {\rm r}_1^2({\it t})d{\rm x}_1^2 -
e^{2x_1}{\Big(}{\rm r}_2^2({\it t})d{\rm x}_2^2 - {\rm r}_3^2({\it t})
d{\rm x}_3^2{\Big)}.$$

For the models considered we have the following structure constants: $\;
{\rm C}^2_{21} = - {\rm C}^2_{12} = {\rm C}^3_{31} = - {\rm C}^3_{13} = 1. \;$
Then, it is easy to find the following relationships:
$${\cal H}^b_a \>= \> 2 {\it diag}({\rm h}_1,\> {\rm h}_2, \> {\rm h}_3);
\quad {\eta} \> = \> {\rm r}^6; \quad {\cal P}^b_a = - {2 \over {{\rm r}_1^2}}
{\delta}^b_a. \eqno(38)$$

By using (38) and the restriction: $\; {\rm j}_{ab} \> = \> 0 \;$ from (22)
and (23) we find:
$${\rm G}^0_0 \; = \; {1 \over 2}{\sum}_{k=1}^3 {\rm h}^2_k - {9 \over 2}
{\rm h}^2 + {3 \over{{\rm r}_1^2}},$$
$${\rm G}^0_1 \; = \; 2{\rm h}_1 - ({\rm h}_2 + {\rm h}_3),$$
$${\rm G}^{\underline a}_{\underline a} \; = \; {\dot {\rm h}_a}
- 3{\dot {\rm h}} + 3{\rm h}{\rm h}_a -  {1 \over 2}{\sum}_{k=1}^3 {\rm h}^2_k
- {9 \over 2}{\rm h}^2 + {1 \over{{\rm r}_1^2}},$$
$${\rm T}^0_{0(eff)} \; = \; (1 - 4\beta\chi - \beta{\rm t}^{\sigma}_{\sigma})
^{-1}{\Big[}{\rho} + {\chi} - {{\beta} \over 4}({\rm t}^{\sigma}_{\sigma}
 + 4\chi)^2{\Big]}
- 6{\it f}_0({1 \over 2}{\dot {\rm M}}{\rm h} +
{1 \over 4}{\dot {\rm M}}^2), $$
$${\rm T}^{\underline a}_{{\underline a}(eff)} \; = \; - (1 - 4\beta\chi
- \beta{\rm t}^{\sigma}_{\sigma})^{-1}{\Big[}{\rm p} - {\chi} +
{{\beta} \over 4}({\rm t}^{\sigma}_{\sigma} + 4\chi)^2{\Big]} -
2{\it f}_0{\Big[}{\ddot {\rm M}} + {\dot {\rm M}}
(3{\rm h} - {\rm h}_a) + {1 \over 4}{\dot {\rm M}}^2{\Big]}. \eqno(39)$$

\noindent It is clear that equation $\; {\rm G}^0_1 \> = \> - {1 \over
{2{\it f}_0}}
{\rm T}^0_{1(eff)} \> = \> 0 \;$ leads to the relationships:
$${\rm h} \; = \; {\rm h}_1 \; = \; {1 \over 2}({\rm h}_2
+ {\rm h}_3). \eqno(40)$$

It follows from (40) that $\; {\rm r}_1 \> = \> {\rm r} \;$. By
virtue of (40) it is easy to show that the equations
$\; {\rm G}^{\underline a}
_{\underline a} = - {1 \over {2{\it f}_0}}{\rm T}^{\underline a}
_{{\underline a}(eff)} \;$ have the following integrals of motion:
$${\rm h}_2 \; = \; {\rm h} + {{\sl s} \over {{\rm r}^3}}(1 - 4\beta\chi
- \beta{\rm t}^{\sigma}_{\sigma})^{-1}; \quad {\rm h}_3 \; = \; {\rm h}
- {{\sl s} \over {{\rm r}^3}}(1 - 4\beta\chi - \beta{\rm t}^{\sigma}
_{\sigma})^{-1}, \eqno(41)$$

\noindent and consequently
$${\rm r}_2 \; = \; {\rm r}{\exp}{\Big(}{\sl s}{\int}{\rm r}^{-3}
(1 - 4\beta\chi - \beta{\rm t}^{\sigma}_{\sigma})^{-1}d{\it t}
{\Big)}; \quad  {\rm r}_3 \; = \; {\rm r}{\exp}{\Big(}{\sl s}{\int}
{\rm r}^{-3}(1 - 4\beta\chi - \beta{\rm t}^{\sigma}_{\sigma})^1d{\it t}
{\Big)}, \eqno(42)$$

\noindent where $\; {\sl s} \;$ is an integration constant and has the sense of
a measure of the anisotropy. By virtue of (40)-(41)
the equation $\; {\rm G}^0_0 = - {1 \over {2{\it f}_0}}{\rm T}^0_{0(eff)} \;$
can be transformed to
$${\Big[}{\rm h} - {\beta \over 2}(1 - 4\beta\chi -
\beta{\rm t}^{\sigma}_{\sigma})^{-1}({\rm t}^{\sigma}_{\sigma})^{.}
{\Big]}^2 \; = \; (1 - 4\beta\chi -
\beta{\rm t}^{\sigma}_{\sigma})^{-1}{\bigg[}{1 \over {6{\it f}_0}}
{\Big(}{\rho} + {\chi} - {\beta \over 4}({\rm t}^{\sigma}_{\sigma}
+ 4\chi)^2{\Big)} +$$
$$+ (1 - 4\beta\chi - \beta{\rm t}^{\sigma}_{\sigma})^{-1}
{{\sl s}^2 \over {3{\rm r}^6}}{\bigg]} + {1 \over {{\rm r}^2}}, \eqno(43)$$

The obtained equation is a direct generalization of the corresponding
equation for open homogeneous isotropic models in MQGT (see [6]), that is
the case, when the measure of the anisotropy $\; {\sl s} \> = \> 0. \;$ In the
given case the "energy-momemntum conservation law" $\; {\tilde \nabla}
_{\mu} {\rm T}^{\lambda \mu}_{eff} \> = \> 0 \;$ has the same form as
eq. (33). In a way similar to that of section 4 we can find solutions
of eq. (43) in an analytic form:
$${\it t - t_0} \; = \; \int_{{\rm r}_0}^{\rm r} {{\Phi}_3({\rm r}) \over
{\Phi}_4^{1/2}({\rm r})}{\rm dr}, \eqno(44)$$

\noindent where
$${\Phi}_3({\rm r}) \; = \; {1 \over {\rm r}}{\bigg[}1 - 4\beta\chi -
\beta{\rm t}^{\sigma}_{\sigma} +{3\beta \over 2}(\rho + {\rm p}){\Big(}
 1 - 3{{\rm dp} \over {d{\rho}}}{\Big)}{\bigg]},$$
$${\Phi}_4({\rm r}) \; = \; {1 \over {6{\it f}_0}}(1 - 4\beta\chi - \beta{\rm
t}
^{\sigma}_{\sigma}){\Big[}{\rho} + \chi - {\beta \over 4}{\big(}{\rm t}
^{\sigma}_{\sigma} + 4\chi{\big)}^2{\Big]}
+ {1 \over {{\rm r}^2}}(1 - 4\beta\chi - \beta{\rm t}
^{\sigma}_{\sigma})^2 + {{{\sl s}^2} \over {3{\rm r}^6}},$$
\section{Discussion}
The exact solutions (37) and (44), that were found in analytic form,
describe dynamics of the spinning-fluid Bianchi type-1 and
perfect-fluid Bianchi type-V respectively. The corresponding
torsion functions have been written in (9). A brief analysis
shows that the given solutions are, in general, "weakly
singular" i.e have following properties:\\
- an energy density is finite at any time: $\; \rho \> \leq \>
{\rho}_{cr} \> < \infty ;$ \\
- a "volume" of "3-space" is positive $\; {\rm V} = {\rm r}^3 =
{\rm (r_1r_2r_3)} \> \geq {\rm V}_{cr} \> > \> 0 \;$ at any time;    \\
- one or two of the three metric functions $\; {\rm r}_k \;$ may become
zero at a finite moment of time when
$\; 1 - 4{\beta \chi} - {\beta}({\rho} - 3{\rm p}) = 0$               \\
- torsion functions diverge $\; {\rm S}({\it t}) = \infty, \;$ when
$\; 1 - 4{\beta \chi} - {\beta}({\rho} - 3{\rm p}) = 0$               \\

However, in some cases (depending on the values of the spin density,
the anisotropy measure, the cosmological constant, the parameter
$\; {\beta} \>$, and the equation of state) we can find truly regular
solutions with finite energy
density, nonzero metric functions and finite torsion functions.
Using methods of qualitative theory of dynamical systems we can
make a detailed analysis of properties of every possible solutions
of the MQGT equations for the models considered.
\acknowledgements
We would like to thank Profs. F. Fedorov and A. Minkevich for
their valuable support during our study in Minsk. N. H. C. would like to
express his gratitude to Prof. A. Zichichi and the World laboratory for
providing the fellowship and Prof. A. Ashtekar for warm hospitality and
encouragement during his stay in Syracuse.

\end{document}